# Probing Electrocatalytic Gas Evolution Reaction at Pt by Force Noise Measurements. Part 2. Oxygen.

Nataraju Bodappa,[1*] Gregory Jerkiewicz,[2] Peter Grutter[1]

[1] Department of Physics, McGill University, 3600 rue University, Montreal, Quebec, Canada H3A 2T8
[2] Department of Chemistry. Queen's University, 90 Bader Lane, Kingston, Ontario, Canada K7L 3N6
*Email: Nataraju.bodappa@mcgill.ca

**ABSTRACT:** Understanding $O_2$ bubble nucleation and growth during the oxygen evolution reaction (OER) is crucial to comprehend their influences on catalytically active sites in the process. To achieve this goal, mapping the spatial variation of nanoscale dynamic individual steps at the electrocatalytic interfaces is vital, as it further enables a detailed understanding of the mechanism of the process. Here, we combined tapping mode AFM imaging with a Pt ultramicroelectrode to investigate oxygen bubble nucleation, growth, and detachment. Our AFM feedback error signal and topography data reveal that bubbles of $O_2$ gas nucleate at the step edge sites and interact with the catalytically active sites. This interaction between primary catalytic sites and bubble nucleation sites is the primary reason for a decrease in the current density at a given high overpotential of the OER. Our findings advance the understanding of the complexity of phenomena involved in gas evolution on catalytic surfaces.

Oxygen ($O_2$) evolution takes place at many different electrocatalysts and requires a high overpotential due to a low exchange current density ($j_o$) of the process.[1] Active sites for many electrocatalytic systems are attributed to defect sites (structural imperfections), such as edges, grain boundaries, and atomic steps [2,3]. During the $O_2$ evolution reaction (OER), the generated supersaturated dissolved $O_2$ at the catalyst surface begins to nucleate and form $O_2$ bubbles. [4,5] The stochastic nature of $O_2$ bubble formation can lead to catalytic active sites being blocked by bubbles, thereby creating an additional ohmic resistance and contributing to an increase in the reaction overpotential. [6,7] Thus, understanding bubble formation on the catalyst's surface is crucial and will enable distinguish between the active sites at which $O_2$ is generated and those at which $O_2$ bubbles are formed.

Formation of individual $O_2$ nanobubbles has previously been studied by electrochemical oxidation of $H_2O_2$ using Pt nanoelectrode voltammetry.[4,5] It was found that the supersaturation ratio ($S = C_g^S/C_g(1\ bar) - 1$) of $O_2$ is ~130 times higher than the equilibrium saturation concentration required to observe $O_2$ nanobubbles on Pt and is independent of electrode radius ~(4-80 nm).[4] Here, $C_g$ is the $O_2$ saturation at the pressure of 1.00 bar and $C_g^S$ corresponds to the supersaturation concentration of dissolved $O_2$ gas required to nucleate the gas phase.

$O_2$ bubbles on a Pt microelectrode were also studied by using high-speed photography in the range of 2.5 V $\leq E \leq$ 4.5 vs the reversible hydrogen electrode (RHE).[8] It was concluded that a small blanket of bubbles with a thickness of ~ 20-25 μm exists. On top of the blanket of small bubbles, large bubbles (~ 100-500 μm) grow through the coalescence of small ones and eventually detach. However, despite this existing knowledge, the initiation of $O_2$ bubble formation at spatial resolution on a microelectrode remains elusive.

Although nanoelectrode voltammetry was successful in estimating the activation barrier, critical saturation concentration, and nucleus size for $O_2$ bubbles, it does not offer spatial resolution. Mapping gas bubble nucleation sites on the macroscopic catalyst surfaces, such as Pt, Au, and $MoS_2$, was carried out using scanning electrochemical cell microscopy (SECCM)[9-11]. In this technique, a microdroplet of electrolyte inside a micropipette tip act as a probe, and the resulting blockage current due to the evolution of a surface nanobubble is studied with spatial mapping. In all these studies, nucleation of such events was noticed to be insensitive to faceting or grain boundaries of the surface.[9,12]

Atomic force microscopy (AFM) has been used to image surface nanobubbles.[13] We recently demonstrated that in-situ AFM enables the investigation of $H_2$ gas bubble formation on a Pt ultramicroelectrode (UME).[14] We observed excess AFM force noise during the $H_2$ evolution reaction, which was related to the formation, growth, and detachment of $H_2$ gas bubbles.[14] Together with current transients, deflection noise, and optical microscopy results, we demonstrated the existence of two different-sized $H_2$ bubbles, related to the bubble blanket, and a coalesced large-size bubble. We measured their detachment dynamics and concluded that coalescence efficiency depends on the geometry of UME.

In this contribution, we analyze current transients and simultaneous cantilever deflection noise at different applied potentials during the OER at a Pt UME in 0.10 M aqueous $H_2SO_4$. Here, our objective is to understand how the nucleation of $O_2$ gas bubbles occurs by combining chronoamperometry and AFM cantilever deflection noise analysis. Then, we applied these insights to acquire *in situ* AFM mapping of sites at which $O_2$ gas bubbles form.

All the experimental methods used in this research are presented and discussed in detail in our previous letter.[14] In particular, we use a three-electrode configuration (Pt UME working electrode, Pt wires as counter and quasi-reference electrodes). The potential of the working electrode was controlled using an external CHI potentiostat. All potential values are reported on the reversible hydrogen electrode (RHE) scale. Conversion of potential values from the Pt quasi-reference electrode scale to the RHE one is discussed in the supporting information (SI). Specifically, the potential of the quasi-reference electrode was determined to be −0.83 V on the RHE scale in the same electrolyte solution (Figure S1). A silicon (Si) cantilever with a spring constant ($k$) of 0.12 N/m was used for deflection noise measurement.

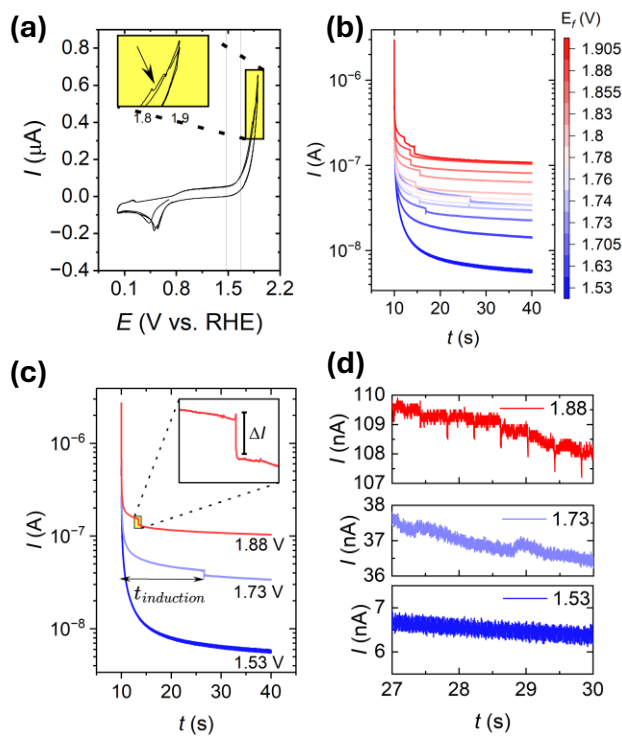

**Figure 1.** (a) Cyclic voltammetry (CV) of Pt UME in a 0.10 M aqueous $H_2SO_4$ solution. Scan rate = 0.1 Vs$^{-1}$ (inset shows the noise in the CV current) (b) Current transient response of the Pt UME in 0.1 M $H_2SO_4$ for different final potential steps ($E_f$) from the open circuit potential. (OCP) (c) Comparison of the steps in the transients (inset shows the spike signatures) (d) Enlarged view of the transient response during OER for different final potential steps.

A cyclic voltammetry (CV) profile of Pt UME in 0.10 M aqueous $H_2SO_4$ acquired at a potential scan rate of 0.10 V s$^{-1}$ is shown in Figure 1a. The CV features of Pt electrodes have been studied for over 50 years and are well understood. They are assigned to the underpotential deposition of H (UPD H) and the surface oxide formation at $E \geq 0.8$ V.[15] In addition, the HER occurs at E < 0.00 V, and a noticeable current due to the OER is observed at $E > 1.53$ V. In this contribution, we focus on the OER regime and observe "noise" in the current when $E \geq \sim 1.85$ V (Inset in Figure 1a).

Figure 1b shows chronoamperometry (CA) transients when the potential is stepped from an initial potential ($E_i$) of 0.83 V to a final potential ($E_f$) that varied from 1.53 to 1.90 V. After applying $E_f$ for 30s, the potential was reversed to $E_i$ for 60s to bring the system to its previous condition, which is an open-circuit potential. The initial high current ($I$) value is due to several concurrent contributions, namely the double-layer charging, the surface oxide formation, and the OER. Subsequently, the value of $I$ decays reaches a steady-state value with the OER being limited by the diffusion of the $O_2$ away from the electrode surface. Once the double layer has been charged, the magnitude of $I$ is associated with the OER because the contribution from the Pt oxide formation is tiny.

Notably, the current versus time ($I$ vs. $t$) transients exhibit no steps or spikes in the case of $E_f \leq 1.70$ V, with the OER being the main contributor to the measured $I$. However, in the case of $E_f \geq 1.705$ V, the $I$ vs. $t$ transients reveal two distinct features: (i) a sharp step of $\Delta I$ (Figures 1c inset) and (ii) spikes (Figure 1d). In particular, current steps in the case of 1.705 V $\leq E_f \leq$ 1.95 V and (ii) spikes in the case of $E_f > 1.83$ (Figure S2). Moreover, for each $E_f$, we observe that one current step is followed by several current spikes. (Figures 1c, d)

We hypothesize that the steps are associated with the formation of gas bubble nuclei and the spikes are related to the detachment of small bubbles followed by the subsequent formation of large bubbles through coalescence. This hypothesis is supported by the $I$ vs. $t$ transients in which the $\Delta I$ steps appeared prior to the appearance of the spikes. The observation of $\Delta I$ steps also supports the heterogeneous nature of the $O_2$ nucleation on the Pt UME.

The time required to observe $\Delta I$ is called the induction time ($t_{induction}$), which represents the time required for the formation of $O_2$ gas nucleus (Figure 1c). A plot of the induction time vs. the applied potential indicates that the value of $t_{induction}$ decreases with increasing applied $E_f$ (Figure 2a). This behavior is expected because the rate of the OER increases with the rising applied potential. (Uncertainty in the measurement is required to discuss the two slopes further). The $\Delta I_{step}$ follows an exponential relationship with the applied overpotential, yielding a slope of 0.25 V/ decade (Figure 2b). These two observations support the interpretation that the steps are due to the $O_2$ nucleation on the surface as a result of potential-controlled activation processes. Note that similar steps and induction times were observed for the nucleation step of metal deposition and gas phase nucleation on nanoelectrodes. [16,17]

Henry's law ($p_{O2} = K_H C_{O2}$; where $K_H$ = 1.3 ×10$^{-3}$ mol L$^{-1}$ atm$^{-1}$) gives the amount of the $O_2$ gas dissolved in the electrolyte at equilibrium. This law relates the amount of the $O_2$ gas dissolved ($C_{O2}$) in the liquid to the partial pressure of $O_2$ gas ($p_{O2}$) above the liquid at constant temperature. For $p_{O2}$= 1.00 atm, the solubility of $O_2$ gas in the electrolyte is 1.3 mM. As the experimental set-up is open and exposed to the atmosphere at room temperature, $p_{O2}$ becomes 0.21 atm (21% $O_2$ in air) and the concentration of dissolved $O_2$ is thus 0.273 mM.

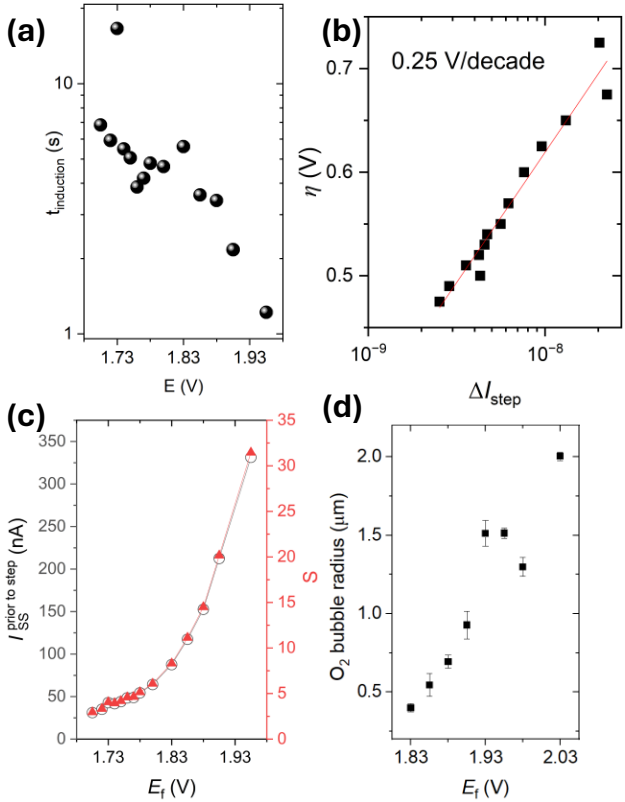

**Figure 2.** (a) Plot of induction time ($t_{induction}$) vs applied potential ($E_f$). (b) Overpotential vs. absolute magnitude of the step responses seen in current transients, $\Delta I_{step}$, on a semi-log plot. (C) Current just before nucleation versus $E_f$ and (d) $O_2$ bubble radius plotted as a function of $E_f$

The steady-state current at the Pt UME equals $I_{SS} = 4D_{O_2}nFr(C_{O_2}^{Surf} - C_{O_2}^{Bulk})$. [4,5] Here, $D_{O_2} = 2 \times 10^{-5}$ cm$^2$ s$^{-1}$ [18,19] is the diffusivity of $O_2$, $n$ is the number of electrons transferred (for OER, $n = 4$), and $r$ is the radius of the UME. The $C_{O_2}^{Surf}$ was estimated by using the $I_{SS}$ just before the $\Delta I$ step in the $I$ vs. $t$ transient (Figure 2c). Because our cell was exposed to air, the supersaturation ratio for $O_2$ ($S = C_g^S/C_g(0.21\ bar) - 1$) was calculated and yielded values in the 3 – 30 range for $I$ varying from 25 to 325 nA.

Integration of the $I$ vs. $t$ transients until the nucleation time yields the total $O_2$ concentration required for gas nucleation to occur and yields $S$ of ca. 30-130 (discussed in *Supporting Information*). The large uncertainty in the reported $S$ values compared to the nanoelectrode measurements is due to the challenges associated with accurately calculating the surface concentration of $O_2$, as the gas exchanges with the atmosphere. However, this is also not a primary objective of the current studies.

The $O_2$ bubble radius was calculated using the total charge integrated under the current spikes using Faraday's law, as described in the supporting information. (Figure 2d, SI). The $O_2$ bubble radius yielded values were found to be in the 0.5 –2 μm range over the ~ 0.2 V potential range. Note that we only observed $O_2$ gas bubbles at the same onset of the potential, ca 1.81 V, using an optical microscope (Figure S3). We observed that the larger size gas bubble radius, optically compared to the electrochemical transients. The observed larger size bubbles formed due to the coalescence of the smaller gas bubbles that detached from the electrode.

AFM deflection noise was monitored simultaneously while collecting the current transients discussed in the previous section. (Figure S3). Figure 3a represents the observed deflection noise for the Si AFM cantilever, which is held with a constant force of 5.5 nN at $E_f$ = 1.53, 1.705, 1.73, and 1.88 V. At $E_f$ = 1.53 V, nearly uniform peak-peak noise (~600 pm) was observed. As the $E_f$ increases to higher potential values (Figure S4), the deflection peak-to-peak noise also increases, and eventually at potentials in the range of 1.705 ≤ $E_f$ ≤ 1.72 V, a specific frequency ~1Hz noise starts to appear (Figure 3a). As $E_f$ increases to 1.73 V, the noise amplitude of the 1 Hz signal increases and saturates with the higher applied potential. These trends can be clearly visualized in the average (10 traces) deflection noise spectral density ($\overline{S_{\delta x}(f)}$) in (Figure 3b) and integrated deflection noise ($\int \overline{S_{\delta x}(f)}$)(Figure 3c). The sharp rise and decay in the deflection noise at 1.73 V are attributed to dynamic interactions of the Si AFM tip with the $O_2$ gas bubble. The sharp rise is primarily attributed to the $O_2$ bubble buoyancy force exerted by the AFM cantilever as it experiences the detachment of an $O_2$ gas bubble from the Pt UME, and the decay is likely due to the depinning/dissolving of the gas bubble and the force feedback resetting exerted by the AFM controller to the constant contact force. Note that the features in the decay traces become different at 1.81 V. This switching is attributed to the large bubble growth via coalescence of the individual $O_2$ gas bubbles.

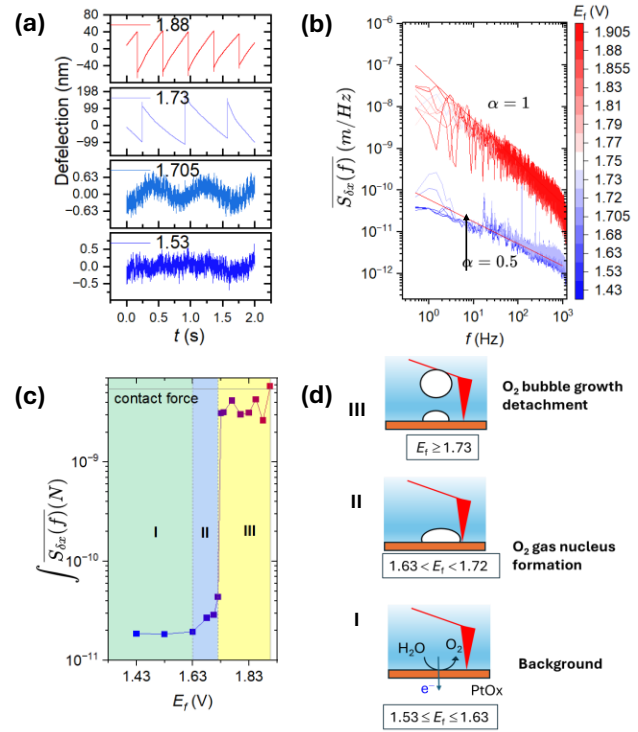

**Figure 3.** *In-situ* deflection noise traces for AFM cantilever at Pt UME-0.1 M $H_2SO_4$ interface at three potentials ($E_f$): 1.53, 1.73, and 1.88 V. Sampling rates: 2 kHz, measurement bandwidth 1 kHz, spring constant of Si-coated force sensor $k$ = 0.12 N/m (PPP-BSI 10). Resolution = 0.5 Hz. The AFM tip was in contact with the sample with a constant force of 5.5 nN and with a very weak feedback gain (0.01). (b) Average deflection noise spectral density for different potential steps. Each spectrum is the average result of 10 spectrum. (c) Semi-log plot for integrated deflection noise versus $E_f$. (d) Schematics of the hypothesized processes at different $E_f$ ranges

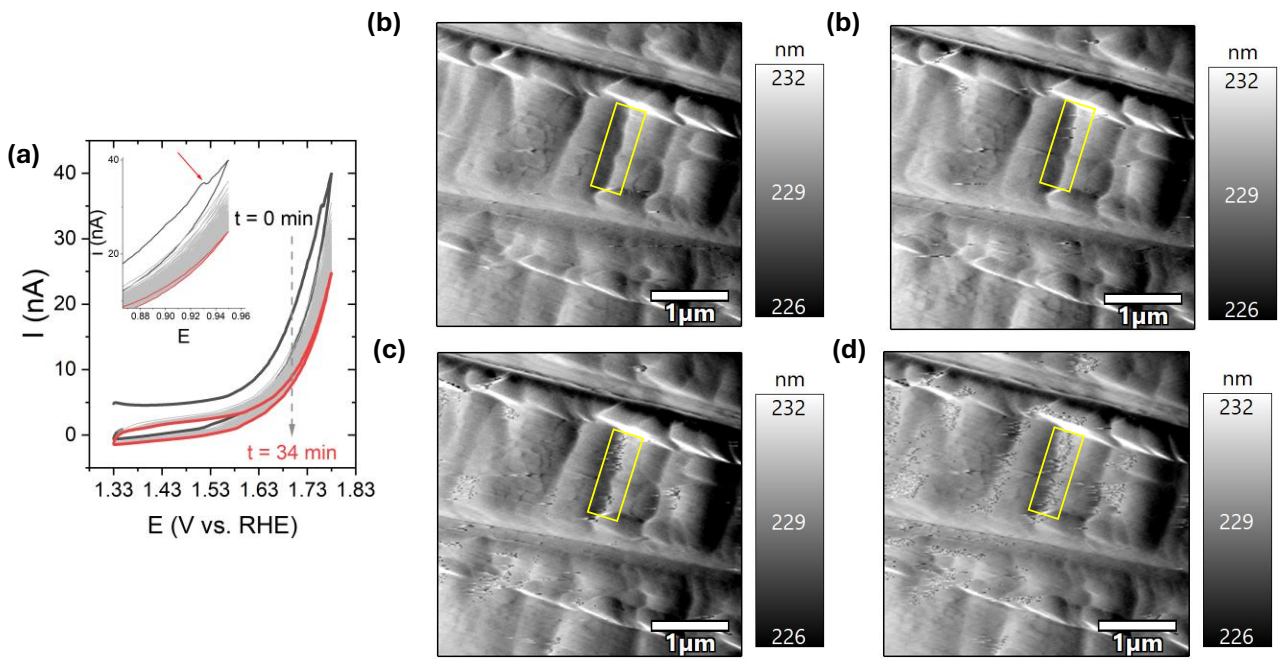

**Figure 4.** (a) The Potential range cycled used to acquire the $O_2$ nucleation mapping by AFM. (b-e) Noise (i.e. AFM amplitude error signal) monitored as a function of time with the applied potential cycling in the range 1.33 V ≤ E ≤1.78 V for 34 min. scan rate of 0.02 V/s. 256×256 lines: 0.5 Hz/line. k = 0.12 N/m

The observed noise as a function of $E_f$ is shown on a semi-log plot (Figure 3c). These regimes are tentatively interpreted as $O_2$ evolution without nucleation, OER with $O_2$ nucleation, and $O_2$ bubble detachment (Figure 3d). Obviously, the resulting noise PSD spectra at 1.43 ≤ $E_f$ ≤ 1.63 V shows only AFM hardware-limited background noise. Note that the onset of OER is 1.53 V. The noise exhibits a clear 1 Hz period with the same background noise within 1.68 ≤ $E_f$ ≤ 1.72 V, range 50 mV. At $E_f$ ≥ 1.73 V, the measured overall noise PSD is close to two orders of magnitude higher than the instrumentation background noise and saturates with the applied potential. The observed saturated noise is due to the resetting of the contact force by the AFM proportional-integral-differential (PID) gains of the AFM controller.

From the data presented in Figures 2d and 3c it is concluded that $O_2$ bubbles are arising from the Pt UME surface starting at $E_f$~ 1.83 V. We assume that the observed saturated deflection noise at lower values 1.73 V ≤ .$E_f$ ≤ 1.83 V arises from smaller $O_2$ bubbles, below the detection limit in current transients.

Acquisition of AFM images for $O_2$ bubbles on Pt UME at a constant $E_f$ was unsuccessful due to the continuous accumulation of $O_2$ at the interface, which causes the AFM tip to lift off from the surface due to the bubble buoyancy force. To avoid this, we employed a potential cycling approach with a limited upper potential ($E_{up}$ = 1.78 V), which is slightly higher than the onset of the saturated force noise detected (~1.73 V) and scan rate of 0.02 Vs$^{-1}$. The CV was continuously cycled between 1.33 V ≤ E ≤ 1.78 V with a rate of 0.02 V/s. (Figure 4a) This procedure enables the $O_2$ formation at the interface and prevents the formation of large $O_2$ bubbles, and thus helps us to identify the nucleation sites for $O_2$ gas bubble formation by AFM imaging.

A large AFM cantilever oscillation amplitude with a feedback setpoint of 85% of the free amplitude) and 0.5 Hz/line scan rate was used for *in-situ* AFM imaging. The error signal of the AFM amplitude scan images (Figure 4b-e) was recorded together with topography (Figure 5) while simultaneously cycling the potential. At the beginning of 8 minutes, barely any bubble noise was seen in the error signal (Figure 4b). However, as time progressed, the noise developed in subsequent scans, localized at the step edges. We thus establish that the initiation of the bubble formation occurs at step edges. (Figure 4c, d, e) The corresponding AFM phase images show strikes along the step edges. (Figure S5) We also noticed the OER catalytic current decreases (Figure 4a) as we see an increase in the noise density with time, further confirming that the bubbles block some of the catalytic sites.

The corresponding surface topography images are shown in Figure 5 The $O_2$ gas bubble height is ~2 nm, which nearly coincides with the $O_2$ gas nucleus size estimated by White et al. using nanoelectrode voltametry during OER.[5]

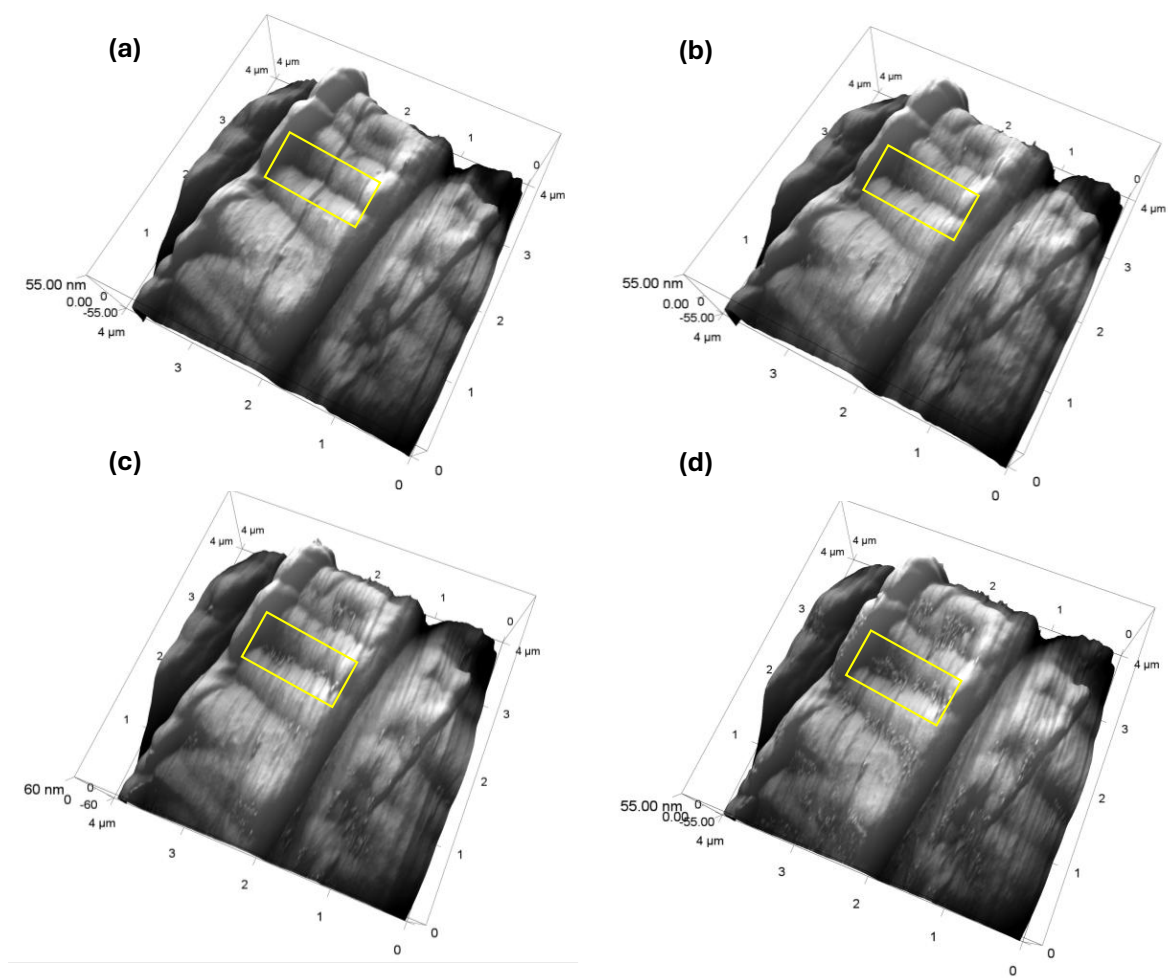

**Figure 5**. (a-d) *3D*-topography scans of Pt UME acquired by *in-situ* AFM during CV cycling. 256×256 lines: 0.5 Hz/line

In conclusion, we have studied AFM tip-sample force noise as a function of applied electrode potential at constant force for the $O_2$ evolution reaction. The excess force noise is related to the $O_2$ gas nucleation, nanobubble growth, detachment, and coalescence growth of $O_2$ gas bubble formation during OER. We have demonstrated proof of concept that tapping mode AFM imaging enables the mapping of the $O_2$ gas bubble nucleation process. The initiation of $O_2$ gas bubble nucleation is favored at step edge sites, thereby blocking catalytically most active sites.

Further. studies to explore higher upper potential during CV cycling enable us to see how the $O_2$ gas bubble nucleation evolves into large bubbles. Extending these studies to well-defined defect sites will help quantitatively understand the bubble interference in electrocatalysis.

## ASSOCIATED CONTENT

**Supporting Information**. Additional data included on the current and force noise during $O_2$ evolution. "This material is available free of charge via the Internet at http://pubs.acs.org." AUTHOR INFORMATION

**Corresponding Author**

* **Nataraju Bodappa -** *Department of Physics, McGill University, Montreal, Quebec, Canada H3A 2T8*.

orcid.org/0000-0003-4387-8887.
Email: Nataraju.bodappa@mcgill.ca.

**Authors**

**Gregory Jerkiewicz** Department of Chemistry. Queen's University, 90 Bader Lane, Kingston, Ontario, Canada K7L 3N6

**Peter Grutter -** Department of Physics, McGill University, Montreal, Quebec, Canada H3A 2T8.

**Author Contributions**

The manuscript was written through the contributions of all authors. All authors have given approval to the final version of the manuscript.

**Notes**

The authors declare no competing financial interest.

## ACKNOWLEDGMENT

This research was supported by FRQNT Teamgrant(FRQ-NT PR-298915), NSERC Discovery Grant, and NRC Collaborative Research and Development (RGPIN-2021-02666 and NRC: CSTIP Grant AI4D-131-1).

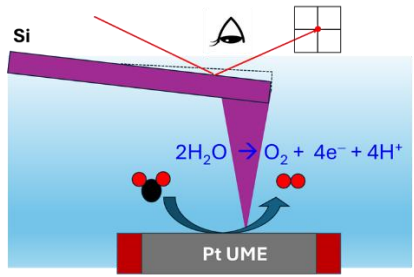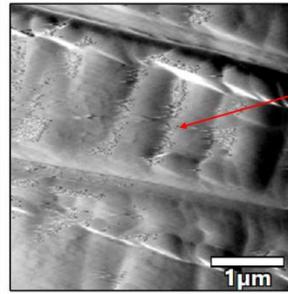

# Probing Electrocatalytic Gas Evolution Reaction at Pt by Force Noise Measurements. Part 2. Oxygen.

Nataraju Bodappa,[1] Gregory Jerkiewicz,[2] Peter Grutter[1]

[1] Department of Physics, McGill University, 3600 rue University, Montreal, Quebec, Canada H3A 2T8
[2] Department of Chemistry. Queen's University, 90 Bader Lane, Kingston, Ontario, Canada K7L 3N6

**1. O$_2$ gas bubble radius**

$$2H_2O \rightarrow O_2 + 4e^- + 4H^+$$

Charge required per mole of O$_2$ formation in the OER = 4F

Number of moles of O$_2$ involved for gas bubble detachment; $n_{O2} = \dfrac{Q}{4F}$     (1)

Here, Q is the total charge under the blip

From the ideal gas equation,

$$n_{O2} = \frac{PV}{RT} = \left(\frac{P_0}{RT}\right)\frac{4}{3}\pi r_b^3 \quad (2)$$

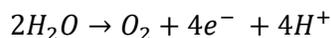

From equations 1 and 2; O$_2$ gas bubble radius is $r_b = \left(\dfrac{3RTQ}{16\pi FP_0}\right)^{1/3}$

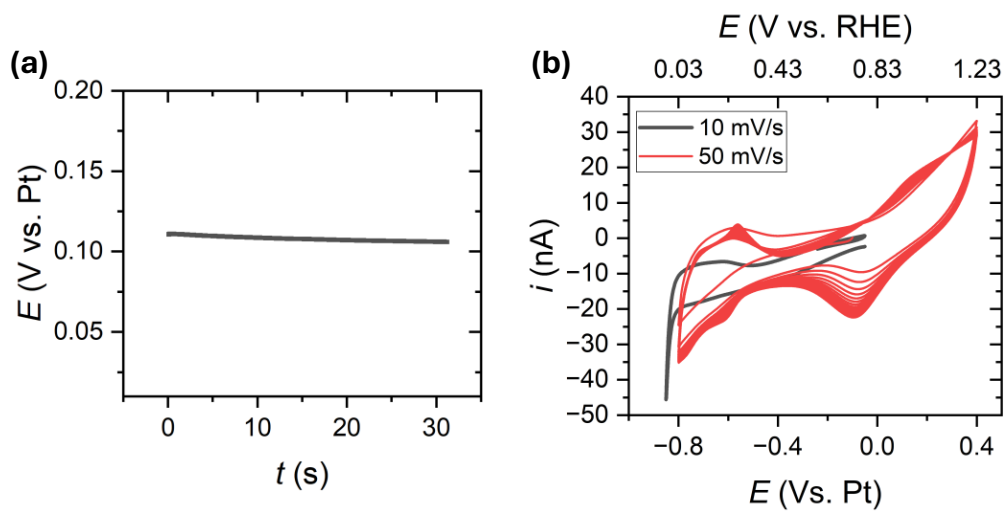

**Figure S1.** (a) Open circuit potential and (b) CV of Pt UME in 0.1 M aerated aqueous M $H_2SO_4$ vs Pt quasi reference electrode.

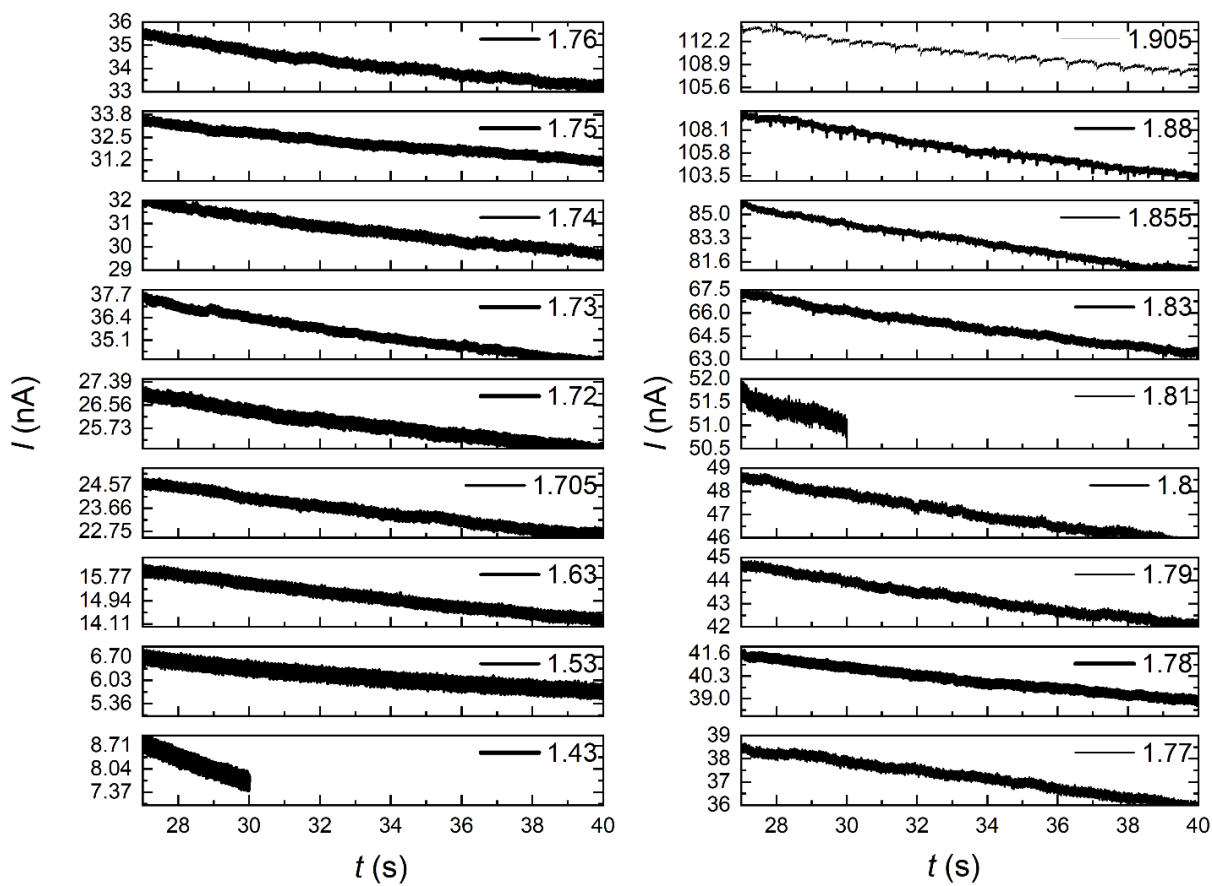

**Figure S2.** Current-time transients in the potential range from 1.43 to 1.905 V

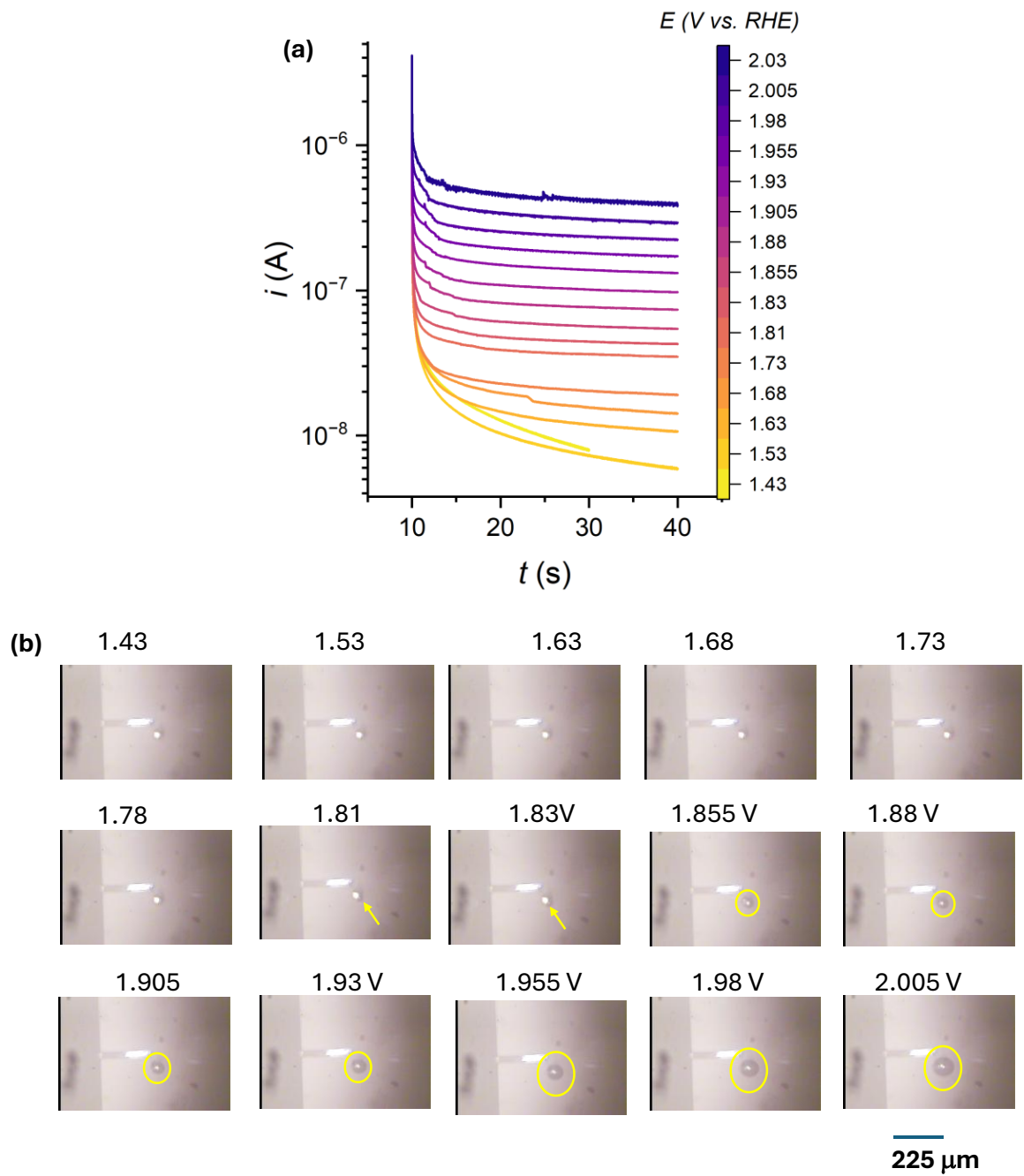

**Figure S3.** (a) Current transients and (b) optical images were simultaneously acquired at each applied potential.

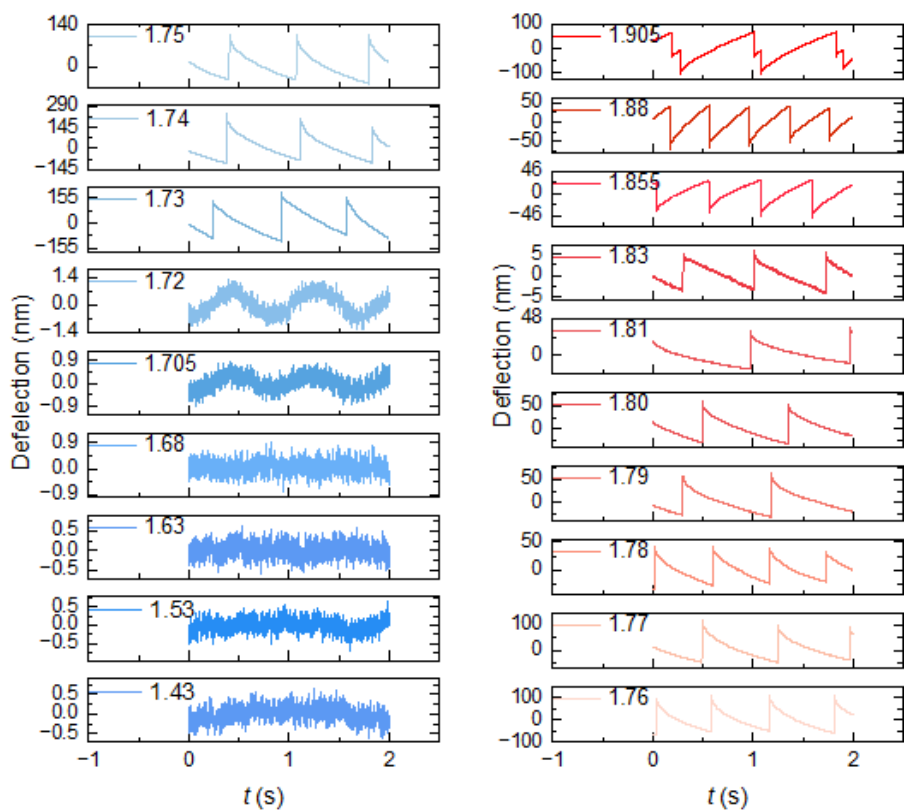

**Figure S4.** *In-situ* deflection noise traces in the potential range from 1.43 to 1.905 V

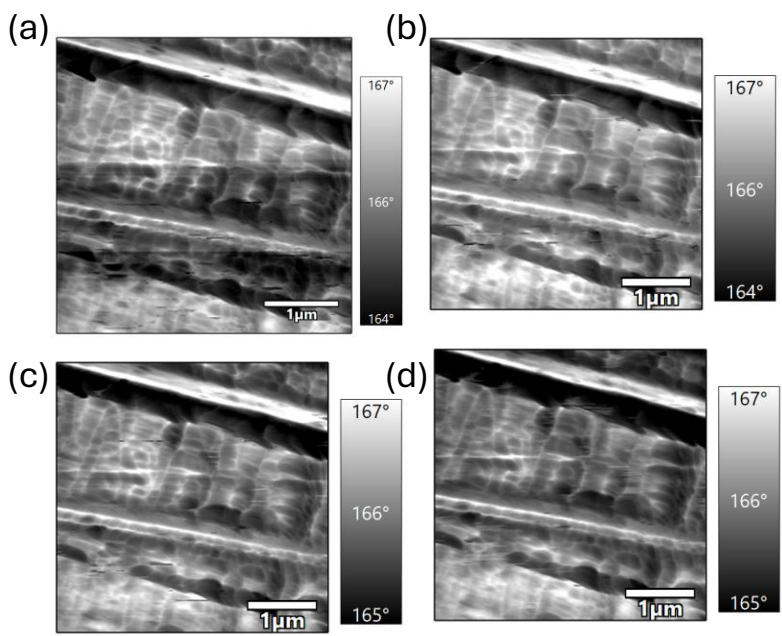

**Figure S5.** *In-situ* AFM phase images of Pt UME during the potential cycling within the range of 1.33 V ≤ E ≤1.78 V for 34 min. scan rate of 0.02 V/s. 256×256 lines: 0.5 Hz/line